\documentclass[10pt, conference, compsocconf]{IEEEtran}
\ifCLASSINFOpdf
   \usepackage[pdftex]{graphicx}
\else
\fi
\usepackage[cmex10]{amsmath}
\usepackage{esvect}

\begin{document}
%
\title{Just-in-Time Memoryless Trust for Crowdsourced IoT Services}


\author{\IEEEauthorblockN{Mohammed Bahutair}
\IEEEauthorblockA{University of Sydney\\
Sydney, Australia\\
mbah6158@uni.sydney.edu.au}
\and
\IEEEauthorblockN{Athman Bouguettaya}
\IEEEauthorblockA{University of Sydney\\
Sydney, Australia\\
athman.bouguettaya@sydney.edu.au}
\and
\IEEEauthorblockN{Azadeh Ghari Neiat}
\IEEEauthorblockA{Deakin University\\
Melbourne, Australia\\
azadeh.gharineiat@deakin.edu.au}
}


%


\maketitle

\begin{abstract}
We propose \emph{just-in-time memoryless trust} for crowdsourced IoT services. We leverage the characteristics of the IoT service environment to evaluate their trustworthiness. A novel framework is devised to assess a service's trust \emph{without} relying on previous knowledge, i.e., \emph{memoryless trust}. The framework exploits \emph{service-session-related} data to offer a trust value valid only during the current session, i.e., \emph{just-in-time trust}. Several experiments are conducted to assess the efficiency of the proposed framework.

\end{abstract}


%
\IEEEpeerreviewmaketitle

\section{Introduction}
Miniaturization and advances in WiFi technologies have led to the emergence of Web-enabled devices. Interconnected devices are typically referred to as the \emph{Internet of Things (IoT)}. More formally, IoT represents the network of \emph{things} that exchange data, essentially enabling a wide variety of applications such as smart homes and smart cities \cite{gubbi2013internet}. A potential application is a crowdsourcing platform whereby IoT devices \emph{crowdsource} services for the benefit of other IoT devices. Crowdsourced services are generally defined as services provided by the crowd to the crowd \cite{brabham2008crowdsourcing}. In \emph{IoT service crowdsourcing}, IoT devices provision services to other nearby IoT devices. We refer to such services as \emph{crowdsourced IoT services}. IoT devices can crowdsource a wide range of service types such as \emph{computing services} \cite{habak2015femto},  \emph{green energy services} \cite{lakhdari2018crowdsourcing}, and \emph{WiFi hotspot services} \cite{Neiat2017caas}. For example, computing services involves IoT devices offering their computational resources (service providers) to other devices. IoT devices with limited computational capabilities (service consumers) may request and consume such services to perform complex tasks that may not be easily achieved otherwise.

Crowdsourcing IoT services pose several trust-related challenges. For instance, let us assume a crowdsourcing environment where \emph{computing services} are provided by IoT devices \cite{habak2015femto}. In such an environment, IoT providers offer their computing resources (CPU, memory) to perform processing tasks for other IoT devices. A potential IoT service consumer may have concerns regarding the service's trust. For example, an untrustworthy service might not protect the privacy of the consumers' data or provide unreliable performance. Similarly, an IoT service provider may have concerns regarding their consumers' trustworthiness. Malicious consumers may misuse IoT services by sending malicious software. Such concerns can be alleviated by ascertaining the trustworthiness of both the service provider and service consumer. IoT environments, however, exhibit certain characteristics that make assessing trust rather challenging. One crucial challenge is the \emph{dynamic nature} of IoT environments. IoT devices are inherently expected to come and go and their existence may not be for long periods. For example, wearables like shirts and shoes have a limited lifespans as people tend to replace them frequently. Additionally, IoT devices can have different owners at any given time. For instance, an IoT shirt can be worn by different people (e.g., bothers). These dynamic characteristics may lead to unreliable historical records that might result in an inaccurate trust assessment. 

\begin{figure*}
    \centering
    \includegraphics[width=0.7\textwidth]{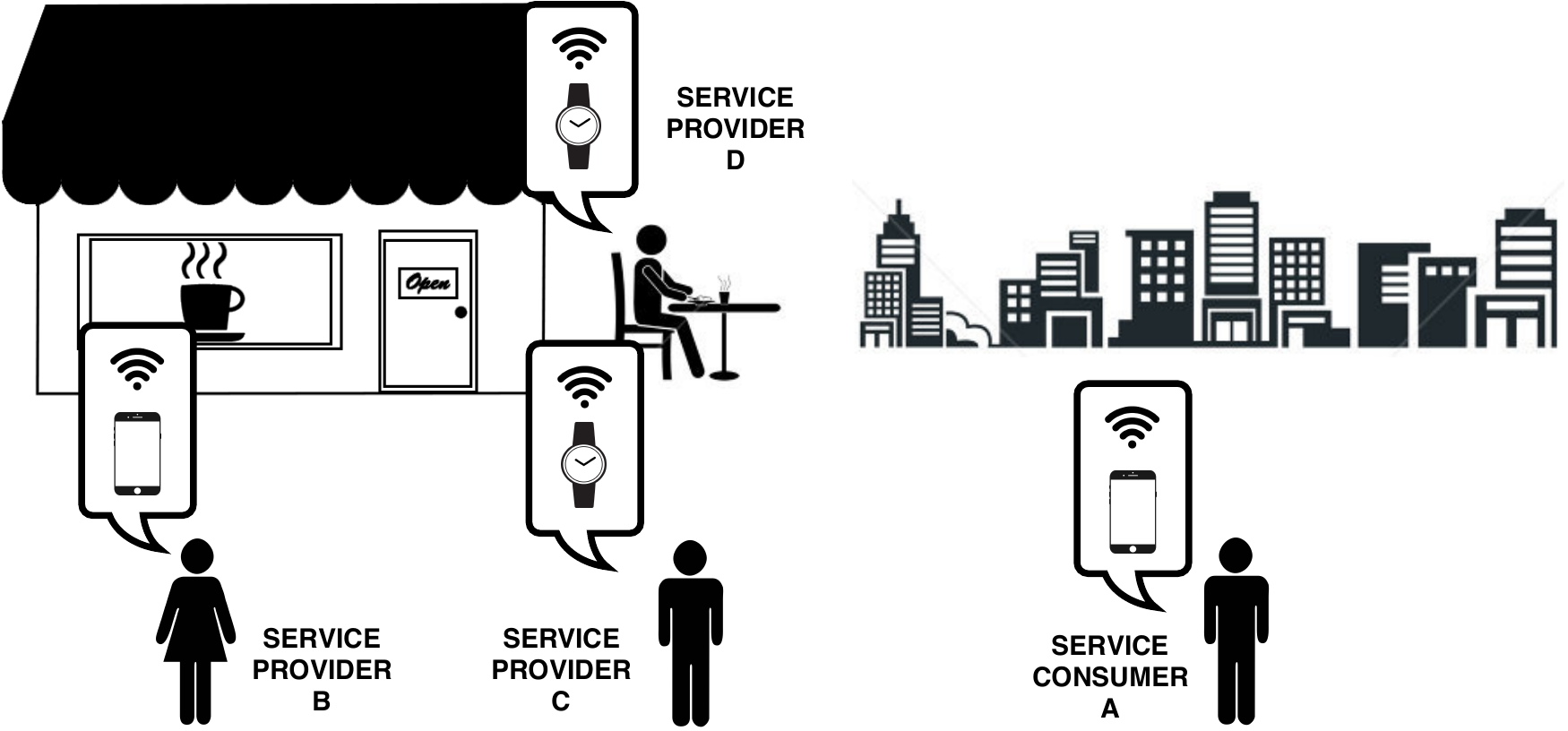}
    \caption{Crowdsourcing WiFi hotspots using IoT devices.}
    \label{fig:motivation_scenario}
\end{figure*}

Previous trust management frameworks relied mainly on historical data to evaluate the trustworthiness of service providers \cite{saied2013trust,chen2011trm}. IoT environments are highly dynamic; i.e., new devices are deployed and removed every day \cite{kyriazis2013smart}. New devices do not generally have prior interactions with other IoT devices. Service consumers, as a result, may not be able to assess the trustworthiness of services provided by new devices. Therefore, the majority of existing trust frameworks cannot be applied to such environments due to the lack of historical records. We introduce the concept of \emph{just-in-time memoryless trust} to overcome these challenges. The trustworthiness of the provider is evaluated \emph{without} relying on historical data (memoryless). We propose to leverage the characteristics of an active IoT service session to compute an accurate trust value for the service. As a result, the assessment would provide an accurate measure of a service's trustworthiness, given a specific service session (just-in-time). We propose a framework that assesses the trustworthiness of IoT services using service-session-related data. Our approach exploits existing IoT devices (bystanders) to assess the trustworthiness of IoT services in their vicinity. We achieve this by using a \emph{collaborative model} where users of the crowdsourcing platform are also expected to participate in assessing the trustworthiness of the services. In that respect, IoT devices that wish to consume IoT services are to contribute to the platform by evaluating IoT services for other consumers. It is worth noting that such a model is used in other popular applications, specifically in peer-to-peer applications, such as Skype and Window 10's Delivery Optimization. For example, Skype users contribute to the platform by acting as relay stations to forward calling packets for other Skype users.

The rest of the paper is organized as follows. Section \ref{section:problem_definition} presents preliminaries and problem definition. Section \ref{section:framework} introduces our just-in-time memoryless trust management framework. Section \ref{section:experiments} discusses the experimental results. Section \ref{section:literature} surveys related work. Section \ref{section:conclusion} concludes the paper and highlights future work.

\subsection*{Motivation Scenario}

We use the following motivation scenario to illustrate the significance of our work (Fig. \ref{fig:motivation_scenario}). Assume a crowdsourcing environment where IoT devices are used to provide/consume services to/from other IoT devices. The provided services in this scenario are WiFi hotspots. In other words, an IoT service provider shares their device's Internet access using an application like WiFiMapper\footnote{https://www.wifimap.io}. Service providers $B$, $C$, and $D$ in Fig. \ref{fig:motivation_scenario} use their smartphones or smartwatches to share their Internet with other nearby IoT devices. Service consumer $A$ wishes to consume one of the available nearby services. 

Service consumer $A$ is presented with the potential service providers $B$, $C$, and $D$. The available providers are unknown to consumer $A$ and, therefore, consumer $A$ has no knowledge about their trustworthiness. Additionally, service providers $B$, $C$, and $D$ do not have historical records that reveal their previous behaviors. Consumer $A$, therefore, fails to ascertain the level of trust from such providers. Service provision/consumption, as a result, does not occur due to the absence of trust. We, therefore, propose a framework that assesses the trustworthiness of the providers \emph{despite the lack of historical records}.

It is worth noting that trust management is not limited to WiFi hotspot services. Other IoT services require trust assurance prior to service consumption. For example, assume an IoT crowdsourcing environment, whereby IoT devices offer their computing resources (e.g., CPU and memory) to other less-capable IoT devices \cite{habak2015femto}. Suppose an IoT device owner $A$ wishes to use their smartphone to provision compute services to other nearby devices. Let us assume that provider $A$ has not provisioned any services before. Provider $A$, therefore, lacks historical data that can describe their performance. Assume a consumer $B$ who has some compute tasks to perform on their confidential data. Provider $A$ is a potential candidate to receive consumer $B$'s task. Consumer $B$, however, may not use the available computing service since provider $A$ is unknown to them. Additionally, Provider $A$ does not have any previous historical records that allow consumer $C$ to deduce a trust value for them.

\begin{figure*}
    \centering
    \includegraphics[width=0.80\textwidth]{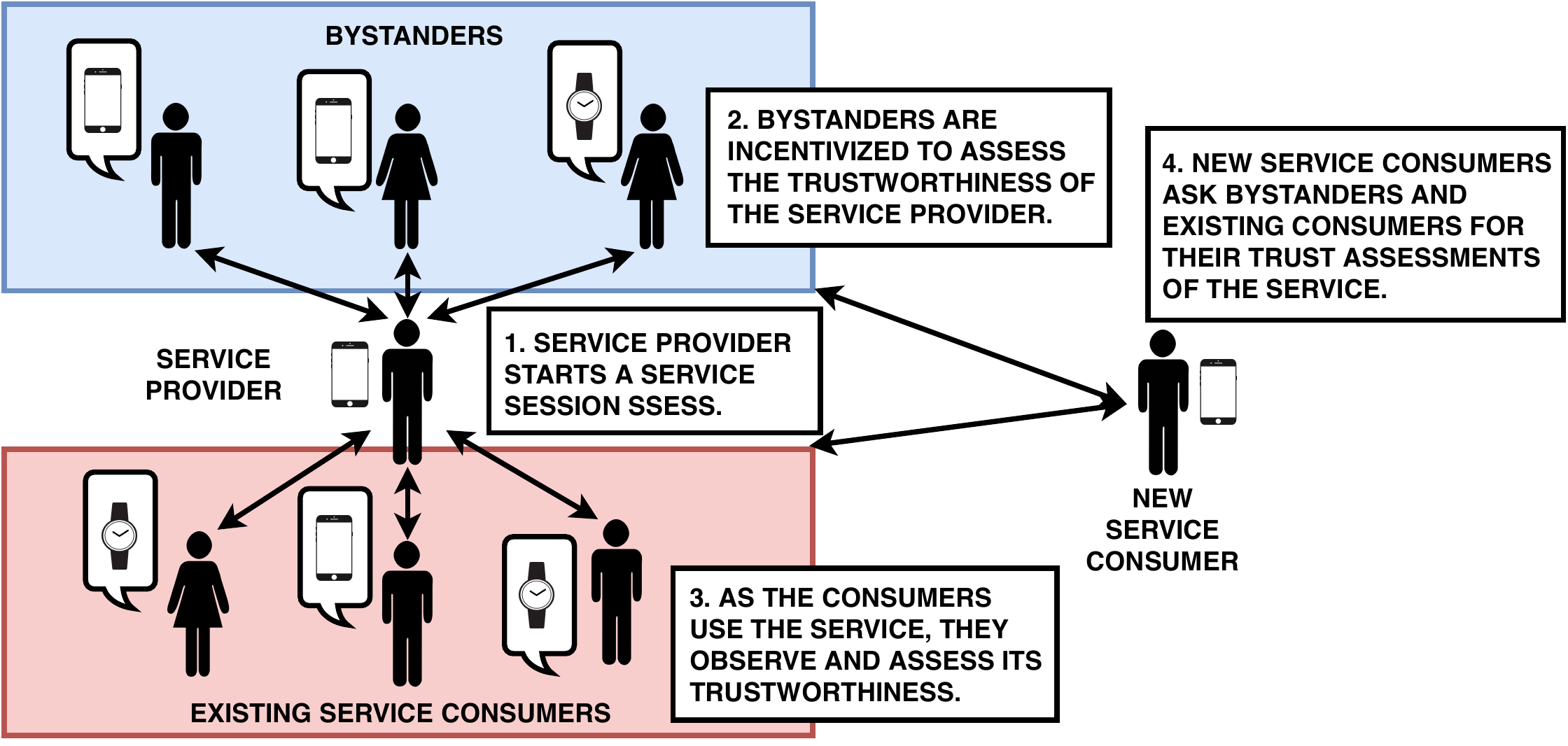}
    \caption{The proposed framework.}
    \label{fig:framework}
\end{figure*}

\section{Preliminaries and Problem Definition}
\label{section:problem_definition}
We model a \emph{crowdsourced IoT service} $S$ provided and consumed by an \emph{IoT device} $d$ as a tuple of  $<id, l, t, d, o, f, q>$ where,

\begin{itemize}
    \item $id$ is a unique service ID. 
    \item $l$ is a GPS location where the service $S$ is provided. 
    \item $t$ is the time interval at which the service $S$ is provided. It is represented as a tuple $<t_s, t_e>$ , where
    \begin{itemize}
        \item  $t_s$ is the start-time of $S$,
        \item  $t_e$ is the end-time of $S$.
    \end{itemize}
        
    \item $d$ is the IoT device that offers the service $S$ (e.g., smartphone).
    
    \item $o$ is the owner of the IoT device. An owner can be an individual or any other cooperation or business (e.g., universities or restaurants).  In this paper, we assume that a service provider is an owner of the IoT device (i.e., $sp = o$).
    
    \item $f$ is a set of functions offered by $S$ (e.g., providing WiFi hotspot). 
    
    \item $q$ is a set of all non-functional parameters of  $S$ (e.g., signal strength). 
\end{itemize}

\label{section:framework}

\subsection{Problem Definition}
As stated earlier, IoT environments are highly dynamic. As a result, the majority of IoT services lack historical records, which are generally used to ascertain their trustworthiness. It is, therefore, crucial to consider other aspects of IoT services to establish their trustworthiness $\mathcal{T}$. The purpose of our work is to identify a function $\mathcal{F}$ that utilizes the properties of a service's environment $\mathcal{E}_S$ to assess its trust. In other words: 

\begin{equation}
    \mathcal{T} \approx \mathcal{F}(\mathcal{E})
\end{equation}

\section{Just-in-Time Memoryless Trust Management Framework}

We introduce a just-in-time memoryless trust management framework for assessing a service's trustworthiness without relying on historical records (see Fig. \ref{fig:framework}). The framework exploits the service session characteristics and its surrounding environment to infer the service's trustworthiness. The framework is divided into three phases: (1) IoT service initiation, (2) IoT service monitoring, and (3) trustworthiness assessment. During the IoT service initiation, an IoT service provider decides to offer their services to nearby devices. Nearby IoT devices monitor the performance of the service during the IoT service monitoring phase. A new consumer relies on existing IoT devices to build a just-in-time memoryless trust value for the potential service provider.  

\subsection{IoT Service Session Initiation}
A service provider $S_p$ starts service provisioning by initiating a \emph{service session} $S_{sess}$. A service session is confined by a geographical area and time duration, e.g., a coffee place between 5:00pm and 7:00pm. Additionally, a service session includes the type of the service being provided, e.g., computing service, WiFi hotspots, etc. The service provider also broadcasts a \emph{promise vector} $\vv{P_S}$ to potential nearby IoT consumers. The promise vector describes the expected behavior of the IoT service. For example, in WiFi hotspot crowdsourcing, a service's trust can be evaluated based on the service's speed, security, and availability.  A possible promise vector can be $[10mbps, Medium, 90\%]$. The vector indicates the service's expected speed is 10mbps, security is medium, and availability is 90\%.

\subsection{Bystanders-Based Trust Assessment}
A service session $S_{sess}$ is limited by space and time. IoT consumers cannot consume the service if they are not within the service session's area and period. For example, IoT consumers at a restaurant cannot consume services from a provider at a coffee shop. Similarly, an IoT service session running between 7:00pm and 8:00pm cannot be consumed between 9:00pm 10:00pm. Many IoT devices may exist within a service's area and periods other than IoT consumers. We refer to such devices as \emph{bystanders}. More formally, we define \emph{IoT service bystanders} as the non-consumer devices that exist within the vicinity of a service session $S_{sess}$ in terms of area and time. 
We leverage IoT service bystanders to assess the trustworthiness of a service $S$ during its session $S_{sess}$. A set of bystanders $B$ receive the promise vector $\vv{P_S}$ from the service provider $S_p$. Additionally, the bystanders assess the service's performance by invoking dummy tasks periodically. For instance, a bystander for a WiFi hotspot service can download a file using a particular service. Each bystander $b \in B$ generates an \emph{observation vector}  $\vv{O_S^b}$ based on their task invocation. The observation vector contains information regarding the actual performance of the service at a specific time. For example, an observation vector $\vv{O_S} = [9mbps, Medium, 80\%]$ for a WiFi hotspot service indicates that the service has a speed of 9mbps, a medium security level, and 80\% availability. An \emph{instantaneous trust} $T_{inst}^b$ is computed by each bystander $b \in B$ using their observation vector $\vv{O_S^b}$ and the IoT service provider's promise vector $\vv{P_S}$. The instantaneous trust reflects the trustworthiness of a particular provider at a time instant from a bystander's perspective.  A bystander $b \in B$ evaluates their instantaneous trust $T_{inst}^b$ using the following equation:

\begin{equation}
    T_{inst}^b = \frac{1}{|\vv{P_S}|}\sum\limits_{i=0}^{|\vv{P_S}|-1} \min\left(1, \frac{\vv{O_S^b}(i)}{\vv{P_S}(i)}\right)
    \label{eq:t_inst}
\end{equation}
where $|\vv{P_S}|$ is the number of elements in the vector $\vv{P_S}$, and $\vv{O_S^b}(i)$ and $\vv{P_S}(i)$ are the $i^{th}$ elements in the vectors $\vv{O_S^b}$ and $\vv{P_S}$, respectively. The value of $T_{inst}^b$ ranges between zero and one, one being highly trusted. For example, assume a WiFi hotspot service provider with a promise vector $\vv{P_S} = [10mbps, Medium, 90\%]$, which represents a service with a speed of 10mbps, Medium security, and 90\% availability. In this example, the security level can either be Low, Medium, or High for simplicity. A representative numerical value can then be given to the security level when used in equation \ref{eq:t_inst} (0 = Low, 1 = Medium, and 2 = High). Additionally, assume a bystander that has an observation vector $\vv{O_S} = [9mbps, High, 80\%]$. The instantaneous trust $T_{inst}$ equals 0.93 after applying equation \ref{eq:t_inst}.


\subsection{Consumer-Based Trust Assessment}
An IoT service provider can offer and provision its services to more than one consumer. We exploit the set of existing consumers $C$ to assess their IoT provider's trust. Similar to IoT service bystanders, IoT consumers receive the provider's promise vector $\vv{P_S}$. Consumers monitor their providers while using the service. Unlike bystanders, consumers can offer a more accurate observation of the IoT service. Consumers perform their monitoring over \emph{periods} rather than single time instances. They, therefore, can capture the behavior of the IoT service over time more accurately and form a better representation for the service's trustworthiness. Each consumer $c \in C$ generates an observation vector  $\vv{O_S^c}$. The observation vector is generated based on the consumer's experience while using the IoT service. The elements of the vector are similar to that of the bystanders' observation vectors; i.e., each represents one aspect of the service's performance. Each consumer $c \in C$ computes the \emph{accumulated trust} $T_{acc}^c$. We define the accumulated trust $T_{acc}^c$ as the trustworthiness of an IoT service from the perspective of the consumer $c \in C$ over their consumption period. Consumers keep updating their accumulated trust as they use the service. The updated accumulated trust utilizes earlier accumulated trust values to increase the accuracy of the service's trust over time. The consumer $c \in C$ evaluates the accumulated trust $T_{acc}^c$ using the following equation:

\begin{equation}
    T_{acc}^c(t+1) = \alpha T_{acc}^c(t) + (1 - \alpha) T_{inst}^c
    \label{eq:t_acc}
\end{equation}
where $T_{acc}^c(t+1)$ is the updated accumulated trust, $T_{acc}^c(t)$ is the current accumulated trust, and $\alpha$ is a weighting factor from zero to one. The $T_{inst}^c$ is the instantaneous trust from consumer $c \in C$'s perspective. The value of $T_{inst}^c$ is computed using equation \ref{eq:t_inst}.

\begin{figure*}
\centering
\begin{minipage}{.47\textwidth}
  \centering
  \includegraphics[width=0.90\linewidth]{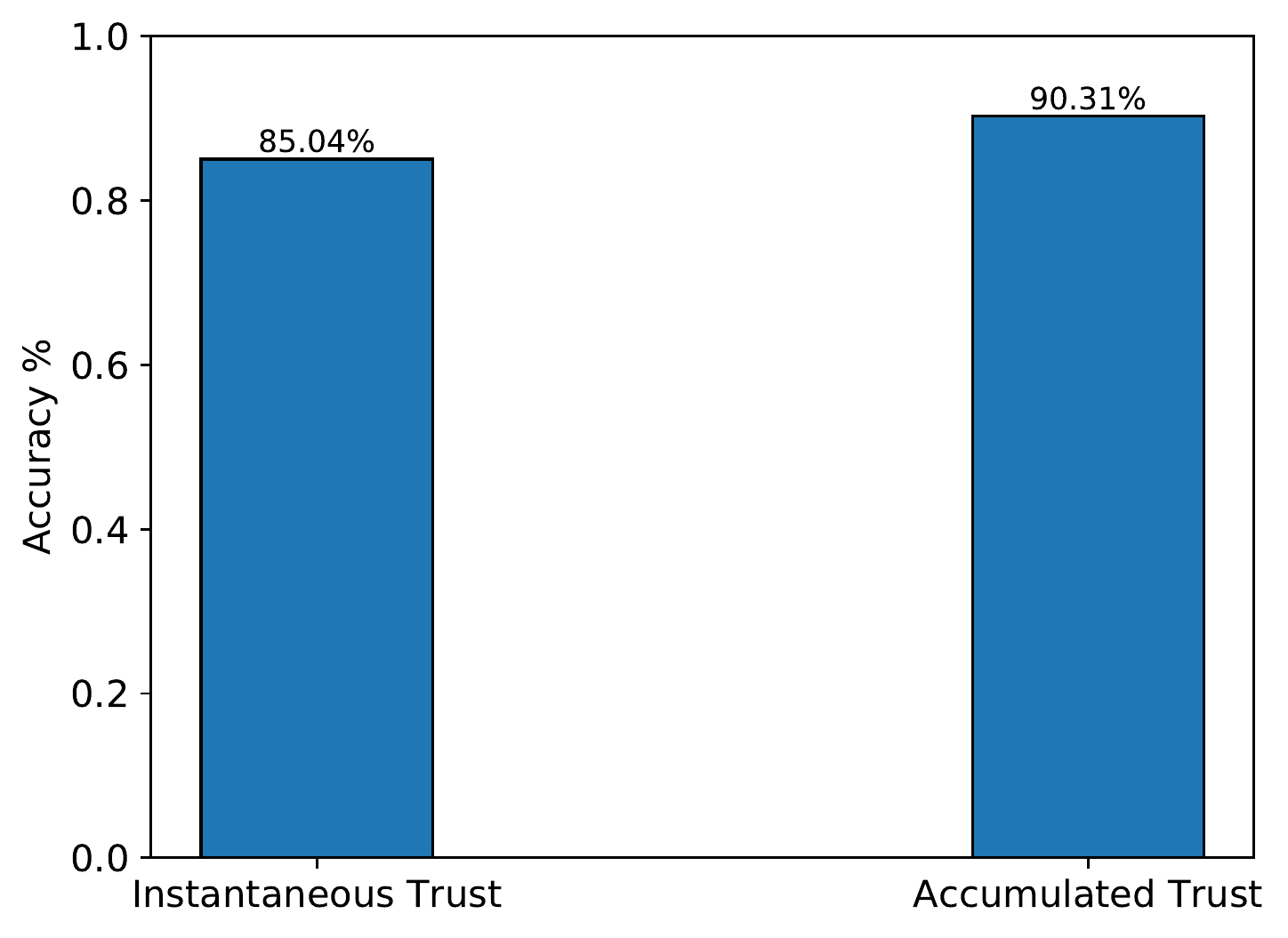}
  \caption{Accuracy in detecting the instantaneous and accumulated trust by existing consumers and bystanders.}
  \label{fig:instantaneous_accumulated_trust}
\end{minipage}%
\hspace{0.5cm}
\begin{minipage}{.47\textwidth}
  \centering
  \includegraphics[width=0.90\linewidth]{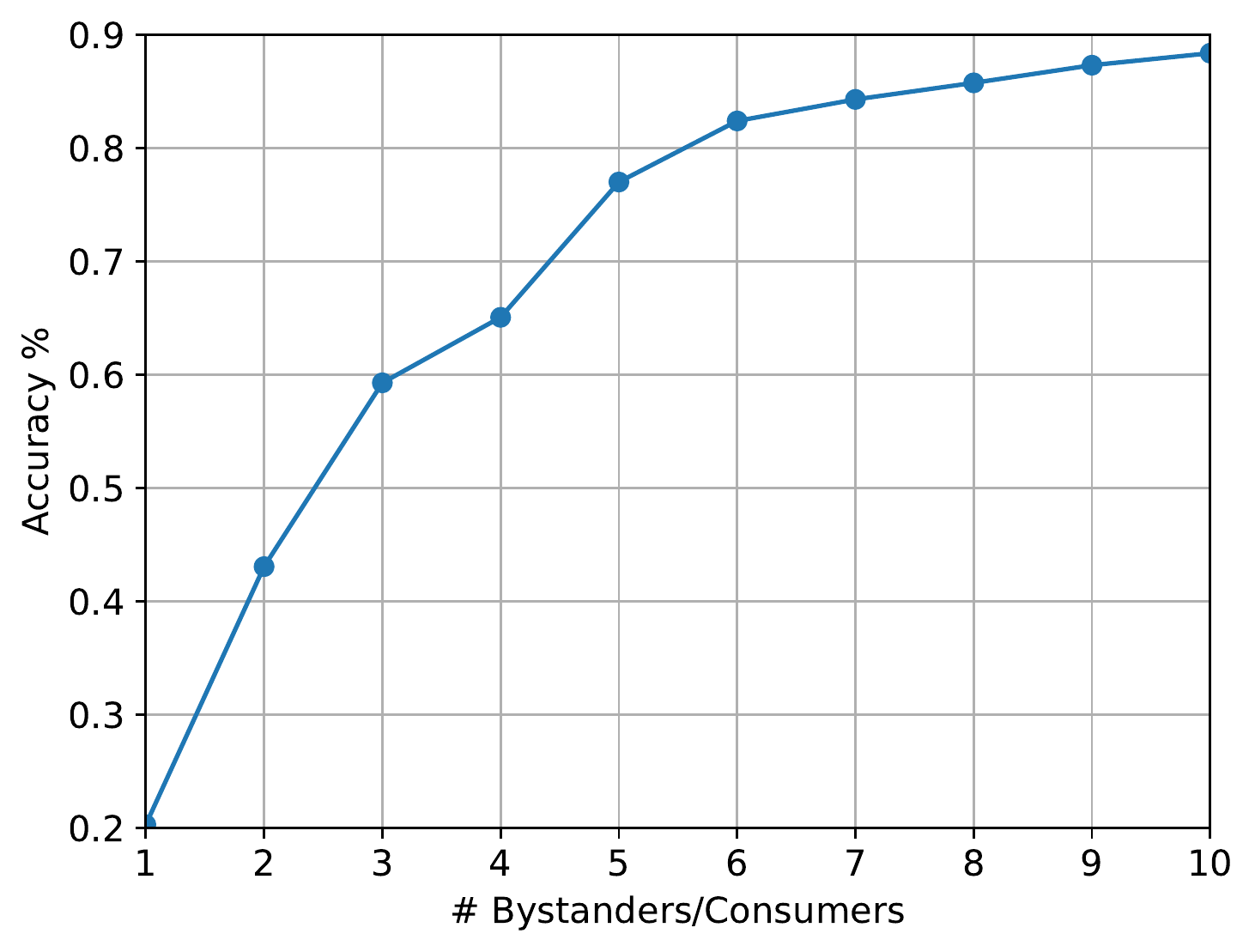}
  \caption{The effect of bystanders/consumers number on the accuracy of the overall trust.}
  \label{fig:bystanders_consumers_effect}
\end{minipage}
\end{figure*}

\subsection{Overall IoT Service Trust Assessment}
A new consumer relies on IoT service bystanders and existing consumers to evaluate an IoT service's trustworthiness.  The accumulated and instantaneous trust values are sent to the new consumer. The new consumer aggregates the trust values to assess the overall trustworthiness of the service using the following equation:
\begin{equation}
    \mathcal{T}_S = \frac{1}{|C| + |B|} \left(\sum\limits_{c \in C} T_{acc}^c + \sum\limits_{b \in B} T_{inst}^b\right)
    \label{eq:trust_basic}
\end{equation}
where $|C|$ is the number of existing consumers, and $|B|$ is the number of bystanders. 

Equation \ref{eq:trust_basic} assesses the trust by weighting instantaneous and accumulated trust values equally. The time at which the instantaneous trust is evaluated should be considered. A \emph{fresher} trust value reflects the service's trustworthiness more accurately. Additionally, the duration that is \emph{covered} by the accumulated trust is crucial. Higher weights should be given to accumulated trust values that cover longer periods. We, therefore, formulate the \emph{freshness} $F_{T_{inst}}$ of the instantaneous trust as follows:
\begin{equation}
    F_{T_{inst}^b} = \frac{ts_b}{TS_B}
    \label{eq:freshness}
\end{equation}
where $ts_b$ is the timestamp at which the instantaneous trust $T_{inst}^b$ is evaluated, and $TS_B$ as the summation of all timestamps from all bystanders in $B$. The freshness value takes a value from zero to one, one indicating a newer trust value. For example, assume three bystanders $A$, $B$, and $C$, which computed the instantaneous trust at different time instances. The three bystanders recorded the timestamp of their observation vectors as an offset from the beginning of the service session. Assume the timestamps are as follows: 2, 8, 20 minutes for bystanders $A$, $B$, and $C$, respectively. The freshness scores for the three bystanders are 0.07, 0.27, 0.67 for bystanders $A$, $B$, and $C$, respectively, according to equation \ref{eq:freshness}. Bystander $C$ has the highest freshness scores since their computed trust is the newest among the three. Similarly, we evaluate the \emph{coverage} $G_{T_{acc}}$ of the accumulated trust as follows:
\begin{equation}
    G_{T_{acc}^c} = \frac{d_c}{D_C}
    \label{eq:coverage}
\end{equation}
where $d_c$ is the duration that is covered by the accumulated trust, and $D_C$ is the summation of all durations covered by all the consumers in $C$. The value of the coverage can take a value between zero and one, where larger values indicate the accumulated trust covered longer periods. For instance, assume a service that is being consumed by three consumers $A$, $B$, and $C$. Each consumer used the service for different periods. Suppose that consumers $A$, $B$, and $C$ used the service for 45, 20, 5 minutes, respectively. The coverage is 0.64, 0.29, and 0.07 for consumers $A$, $B$, and $C$, respectively. Consumer $A$ is awarded the highest coverage since they used the service the longest.

We use the freshness and coverage measures in equation \ref{eq:trust_basic} as follows:
\begin{equation}
    \mathcal{T}_S = \beta \sum\limits_{c \in C} G_{T_{acc}^c} T_{acc}^c + (1 - \beta) \sum\limits_{b \in B} F_{T_{inst}^b} T_{inst}^b
    \label{eq:trust_freshness_coverage}
\end{equation}
where $\beta$ is a user-defined weighting factor between zero and one.

In some scenarios, a bystander/consumer may report inaccurate or wrong trust assessments to the new consumer (e.g., biased or malicious bystander). We introduce the \emph{credibility} measure to overcome this issue. The basic intuition is that biased/malicious bystanders/consumers are special cases. In other words, the majority of consumers/bystanders behave as expected, while only a subset may produce invalid assessments. We, therefore, compare the trust assessments from different bystanders and consumers. The credibility of each bystander/consumer can then be inferred from such comparison. For example, assume, for simplicity, four bystanders with four instantaneous trust values: 0.9, 0.85, 1.0, and 0.1. The first three trust values are close to each other, while the last is significantly different. It can be deduced that the forth bystander may have invalid trust assessment since the majority produced different values. We, therefore, evaluate the credibility of a bystander/consumer $\mathcal{C}_{b|c}$ as follows:
\begin{equation}
    \mathcal{C}_{b|c} = 1 - |T - T_{avg}|
    \label{eq:credibility}
\end{equation}
where $T$ is the instantaneous or the accumulated trust and $T_{avg}$ is the average of all trust values from bystanders and consumers. The credibility value takes a value between zero and one, one indicating high credibility.

Using our earlier example, the credibilities for the four bystanders using equation \ref{eq:credibility} are 0.8125, 0.8625, 0.7125, and 0.3875. The first three bystanders got relatively high scores since they provide similar assessments. The fourth bystander is given the lowest since its trust assessment deviates from the majority.

We use the credibility of the bystanders/consumers by applying equation \ref{eq:credibility} into equation \ref{eq:trust_freshness_coverage} as follows:
\begin{equation}
    \mathcal{T}_S = \beta \sum\limits_{c \in C} \mathcal{C}_c G_{T_{acc}^c} T_{acc}^c + (1 - \beta) \sum\limits_{b \in B} \mathcal{C}_b F_{T_{inst}^b} T_{inst}^b
    \label{eq:trust}
\end{equation}

\begin{figure*}
\centering
\begin{minipage}{.47\textwidth}
  \centering
  \includegraphics[width=0.90\linewidth]{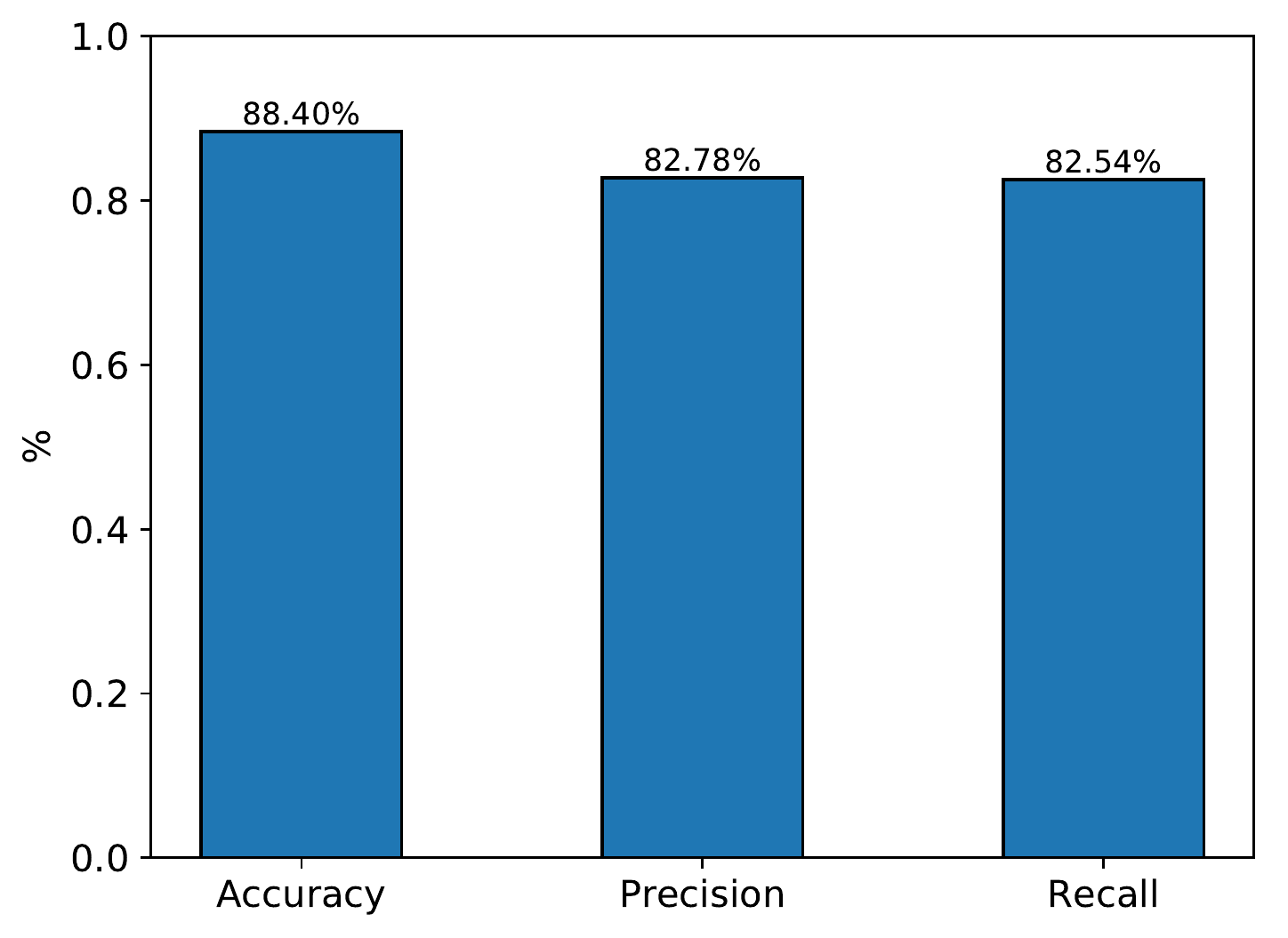}
  \caption{Accuracy, Precision, and recall for the proposed framework.}
  \label{fig:experiment_accuracy}
\end{minipage}%
\hspace{0.5cm}
\begin{minipage}{.47\textwidth}
  \centering
  \includegraphics[width=0.90\linewidth]{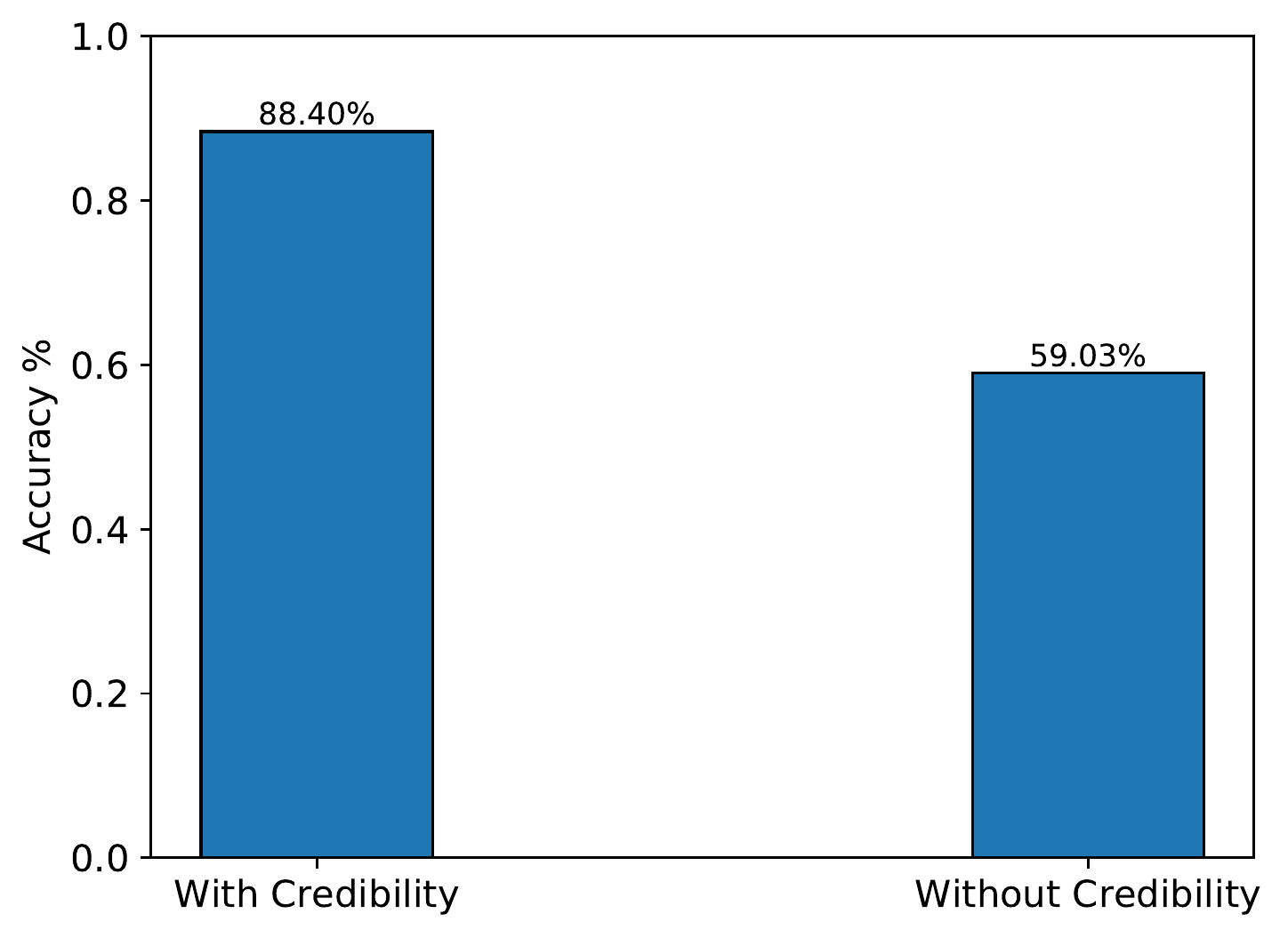}
  \caption{The effect of credibility on the trust accuracy.}
  \label{fig:credibility_effect}
\end{minipage}
\end{figure*}

\section{Experiments}
\label{section:experiments}
We evaluate the effectiveness of the framework in terms of accuracy. More specifically, we examine the accuracy of the just-in-time memoryless trust for IoT services. Additionally, we evaluate the accuracy of the instantaneous and accumulated trust discussed in Section \ref{section:framework}. We run the experiments on a 3.60GHz Intel(R) Core(TM) i7-7700 and 8 GB of RAM.

\subsection{Dataset Description}
Amazon Mechanical Turk\footnote{https://www.mturk.com/} (MTruk) is used to collect the dataset for our experiments. We divided MTurk workers into three categories: bystanders, existing consumers, and new consumers. Workers in all categories are asked to assume that they are at a public place, e.g., a restaurant, where other nearby users are sharing their Internet via WiFi hotspots using smartphones and smartwatches. In such an environment, potential consumers can use an application on their smartphones/smartwatches to get a list of nearby WiFi hotspots. Each provider in the application sets its expected performance, i.e., promise vector. Each worker is presented with a specific WiFi hotspot, with its promise vector. 

Workers in the bystanders' category are asked to assume that they are given some monetary rewards to monitor the performance of service providers. Monitoring is carried out by using the service at specific time instances and recording their actual performance, i.e., observation vector. The performance of the service at different timestamps is given to the workers. The workers are asked to rate the service based on the given performance.

Existing consumes category consists of workers that are treated as current consumers for the presented service. Similar to bystanders category, workers in this category are presented with a list of performance records of the service, i.e., observation vector. Each performance record represents the performance of the service during a period rather than at timestamps. The workers are asked to rate the service based on the given performance. 

Workers in the new consumers' category are those who wish to compute the just-in-time memoryless trust. Each worker is presented with two lists of ratings. The first list represents the service's ratings from consumers who use it. The second list contains bystanders who monitor the service and report their ratings. The workers are asked to consider the ratings from both lists and give an expected rate for the service. 

All workers are asked to give their rating for the services as a value between zero to ten, ten being the highest. The data has been collected using the results from a total of 5000 workers for each category.

\subsection{Experimental Results}
We use \emph{precision}, \emph{recall}, and \emph{accuracy} \cite{Olson2008ADM1795943} to determine the efficiency of our work. We assume that trust can be at one of three levels: \emph{highly trusted}, \emph{moderately trusted}, and \emph{lowly trusted}. Given a trust level $l$, e.g., moderately trusted, \emph{precision} $l$ is defined as the ratio between the number of samples correctly detected as $l$ to the total number of samples detected as $l$:

\begin{equation}
    \label{eq:precision}
    Precision_l =\frac{|correct_l|}{|detected_l|}
\end{equation}

\emph{Recall} for $l$ is the ratio between the correctly detected samples as $l$ to the total number of actual $l$ samples:

\begin{equation}
    \label{eq:recall}
    Recall_l =\frac{|correct_l|}{|actual_l|}
\end{equation}

The \emph{accuracy} in detecting $l$ is the ratio between the number of correctly detected samples as $l$ and the number of correctly detected samples as not $l$ to the total samples number:

\begin{equation}
    \label{eq:accuracy}
    Accuracy_l =\frac{|correct_l|+|correct\_not_l|}{|samples|}
\end{equation}

The first experiment set evaluates the overall accuracy of the whole framework, see Fig. \ref{fig:experiment_accuracy}. The framework achieves high accuracy, precision, and recall scores: 88.40\%, 82.78\%, and 82.54\%, respectively. The second experiment evaluates the credibility measure influence on the accuracy scores. Fig. \ref{fig:credibility_effect} shows the results for this experiment. The framework scores higher in terms of accuracy (around 88.40\%) when the credibility measure is used. The accuracy of the framework reduces significantly when the credibility of the bystander/consumer is not considered (around 59.03\%). Recall that the credibility measures the truthfulness of the trust values provided by the bystanders/consumers. The measure is used to weight the significance of the reported trust from the bystanders/consumers. The absence of the credibility results in treating all reported trust values equally. As a result, biased/malicious bystanders and consumers can greatly disrupt the accuracy of the assessed trust as evident in Fig. \ref{eq:credibility}.

The overall trust value of the service is computed based on the instantaneous and accumulated trust. We evaluate the accuracy in assessing such trust values in the next experiment (see Fig. \ref{fig:instantaneous_accumulated_trust}. The accuracy in computing the instantaneous trust is around 85.04\%, whereas a higher accuracy is achieved when evaluating the accumulated trust; around 90.31\%. The accumulated trust has higher accuracy due to how it is computed. The instantaneous trust is computed from a single observation vector given a specific timestamp. However,  computing the accumulated trust utilizes previously computed instantaneous trust values to get a more accurate assessment for the service.


The last experiment studies the effect of the number of bystanders/consumers on the overall trust of the service. Fig. \ref{fig:bystanders_consumers_effect} shows the results for this experiment. The accuracy of the framework has significantly lower scores when a small number of bystanders and consumers is considered. For example, the accuracy is around 20\% when only one bystander or consumer is used. The accuracy increases exponentially as the number of bystanders and consumers increases. High accuracy scores are reached when six bystanders/consumers are considered (about 82\%). A lower number of bystanders/consumers might not represent the actual service trustworthiness accurately. For example, biased/malicious bystanders/consumers have a greater influence when the total number of bystanders/consumer is low. As a result, the computed trust value does not accurately represent the service's actual trustworthiness.

\section{Related Work}
\label{section:literature}
Trust assessment in crowdsourced IoT service environment is fairly new. Most of the proposed approaches can be grouped into two main categories: \emph{previous experiences-based trust evaluation}, and \emph{social networks-based trust evaluation}.

\subsection{Measuring Trust Using Previous Experiences}
A centralized trust management system (TMS) is proposed to evaluate the trustworthiness of an IoT device based on its past behavior \cite{saied2013trust}. The TMS has four phases: information gathering, entity selection, transaction and evaluation, and learning. The information gathering phase involves collecting data about the executed services by the IoT device. In the entity selection phase, the TMS returns the most trustworthy IoT device for requested services. Performing the task by the IoT service provider happens at the transaction and evaluation phase. The requester gives a score to the provider based on the executed service outcome. In the learning phase, the TMS learns the credibility of the requesters to weight their scores. While TMS performed well according to the conducted experiments, their choice of a centralized solution may act as a bottleneck. 


A trust model is proposed for evaluating the trustworthiness and reputation of nodes in Wireless Sensor Networks (WSNs) \cite{chen2011trm}. The node's reputation is computed based on its performance characteristics: packet delivery, forwarding ratio, and energy consumption. The reputation is later used to evaluate the trustworthiness of the node. The proposed model uses WSN-specific characteristics (e.g., packet delivery which makes it unsuitable for other applications). 

Social IoT network is used to measure the trust between two IoT devices in \cite{nitti2012subjective}. A social IoT network is a type of social networks where nodes are IoT devices. Relationships between IoT devices indicate one or more of the following relations: similar owner, co-location, co-work, social relation, or brand. Each node computes the trust of its friends. The trust is measured based on the individual's and friends' previous experiences. A trust model is also proposed in \cite{wang2016toward} that uses social IoT to manage the interactions between IoT service consumers and providers. The social IoT network is used to search for candidate service providers. Reputation-based trustworthiness is used to evaluate the service providers using previous interactions.

A trust management protocol is proposed in \cite{bao2012dynamic} \cite{bao2012trust} to assess the trustworthiness of IoT devices based on honesty, cooperativeness, and community-interest. Honesty is measured using the direct observation of an IoT device (high recommendation discrepancy, delays, etc.). Cooperativeness and community-interest are computed using data from social networks. Common friends between two IoT owners indicate high cooperativeness between their IoT devices. Community-interest depends on the common communities number between two IoT owners.
A framework is proposed for crowdsourcing services to IoT devices based on their mobility and trustworthiness \cite{kantarci2014mobility}. A central authority exists to manage interactions between the consumer and provider. The trustworthiness of a service provider is computed based on their reputation. Basically, when the central authority receives a task request from the consumer, the task is submitted to multiple service providers. The server then computes the anomalies among the results from the service providers. Service providers with deviated results are marked and their reputation is decreased.

Aforementioned approaches use historical data (i.e., previous experiences) to asses the trust. On the other hand, one key characteristic of crowdsourced IoT service environments is their high dynamism in terms of IoT devices deployment. Every day a large number of IoT services are being added. Newly added services do not have previous records. If those previous experiences are missing, the evaluated trust cannot be accurate. Therefore, these approaches cannot be utilized to accurately measure IoT services' trust.

\subsection{Measuring Trust Using Social Networks}

A social compute cloud framework is proposed in \cite{Caton2014} where users in a social network can share and consume services from other users. The framework leverages the social structure of the network. Relation types between users (e.g., family, colleagues, etc.) are also utilized to determine the level of trust between them. 
A framework is proposed in \cite{Cao2015} that aims at eliminating the privacy risks accompanied with public WiFi hotspots. Social WiFi utilizes social networks relationships to match hotspot users to trusted hotspot providers. The proposed framework lacks generality as it can only be used for WiFi hotspot services.

An approach for evaluating the trust between users in social networks is presented in \cite{adali2010measuring}. Behavioral interactions are used to indicate the level of trust (i.e., conversations between users and message propagation). A conversation between two users can indicate a higher level of trust if: (1) it happens many times, (2) it lasts for a long duration, and (3) there is a balanced contribution of messages from both users. The message propagation indicates the willingness of a user $B$ to forward a message received by another user $A$. A large number of forwarded messages reflect a higher trust value for the sender.
The above work focuses on social networks' relationships which is not sufficient to evaluate the trust between the IoT service provider and consumer. For example, two friends on a social network do not necessitate mutual trust between them \cite{sherchan2013survey}.

\section{Conclusion}
\label{section:conclusion}
We presented just-in-time memoryless trust, a trust evaluation specifically suited for IoT services. The just-in-time memoryless trust accounts for the high dynamism exhibited in IoT environments. Such a dynamic nature causes the lack of \emph{historical records}, a crucial cornerstone for computing the trustworthiness in the majority of existing trust management frameworks. A novel framework was proposed to measure IoT services' trustworthiness \emph{without} relying on previous historical data. The framework exploits session-related data to assess the trustworthiness of IoT services. More specifically, we leverage the experience of IoT bystanders and consumers to build an accurate trust value for a given IoT service. The proposed framework achieved high accuracy scores in our experiments. Future directions can be investigating the trustworthiness of the IoT service consumer.



\bibliographystyle{IEEEtran}
\bibliography{IEEEabrv, ./ref.bib}
%



\end{document}